\documentclass[superscriptaddress,reprint]{revtex4-2} 

\usepackage[utf8]{inputenc}
\usepackage{bm}
\usepackage{graphicx}

\usepackage{bbold}
\usepackage{color}
\usepackage{hyperref}
\hypersetup{colorlinks=true}
\usepackage{tikz}
\usepackage{float}
\usepackage{amsmath}
\usepackage{ulem}
\usepackage{braket}

\newcommand{\refsec}[1]{Section~\ref{#1}}
\newcommand{\refeq}[1]{Eq.~(\ref{#1})}

\begin{document}

\title{Modelling noise in global M{\o}lmer-S{\o}rensen interactions applied to quantum approximate optimization}

\author{Phillip C. Lotshaw}
\email[]{lotshawpc@ornl.gov}
\affiliation{Quantum Information Science Section, Oak Ridge National Laboratory, Oak Ridge, TN 37381, USA}
\thanks{This manuscript has been authored by UT-Battelle, LLC, under Contract No. DE-AC0500OR22725 with the U.S. Department of Energy. The United States Government retains and the publisher, by accepting the article for publication, acknowledges that the United States Government retains a non-exclusive, paid-up, irrevocable, world-wide license to publish or reproduce the published form of this manuscript, or allow others to do so, for the United States Government purposes. The Department of Energy will provide public access to these results of federally sponsored research in accordance with the DOE Public Access Plan.}

\author{Kevin D. Battles}
\affiliation{Georgia Tech Research Institute, Atlanta, GA 30332, USA}

\author{Bryan Gard}
\affiliation{Georgia Tech Research Institute, Atlanta, GA 30332, USA}

\author{Gilles Buchs}
\affiliation{Quantum Information Science Section, Oak Ridge National Laboratory, Oak Ridge, TN 37381, USA}
\affiliation{Quantum Science Center, Oak Ridge National Laboratory, Oak Ridge, TN 37381, USA}

\author{Travis S. Humble}
\affiliation{Quantum Information Science Section, Oak Ridge National Laboratory, Oak Ridge, TN 37381, USA}
\affiliation{Quantum Science Center, Oak Ridge National Laboratory, Oak Ridge, TN 37381, USA}

\author{Creston D. Herold}
\email[]{Creston.Herold@gtri.gatech.edu}
\affiliation{Georgia Tech Research Institute, Atlanta, GA 30332, USA}

\begin{abstract}
Many-qubit M{\o}lmer-S{\o}rensen (MS) interactions applied to trapped ions offer unique capabilities for quantum information processing, with applications including quantum simulation and the quantum approximate optimization algorithm (QAOA).  Here, we develop a physical model to describe many-qubit MS interactions under four sources of experimental noise: vibrational mode frequency fluctuations, laser power fluctuations, thermal initial vibrational states, and state preparation and measurement errors. The model parameterizes these errors from simple experimental measurements,  without free parameters.  We validate the model in comparison with experiments that implement sequences of MS interactions on two $^{171}$Yb$^+$ ions. The model shows reasonable agreement after several MS interactions as quantified by the  reduced chi-squared statistic $\chi^2_\mathrm{red} \approx 2$. As an application we examine MaxCut QAOA experiments on three and six ions. The experimental performance is quantified by approximation ratios that are $91\%$ and $83\%$ of the optimal theoretical values. Our model predicts $0.93^{+0.03}_{-0.02}$ and $0.95^{+0.04}_{-0.03}$, respectively, with disagreement in the latter value attributable to secondary noise sources beyond those considered in our analysis. With realistic experimental improvements to reduce measurement error and radial trap frequency variations the model achieves approximation ratios that are 99$\%$ of the optimal. Incorporating these improvements into future experiments is expected to reveal new aspects of noise for future modeling and experimental improvements.
\end{abstract}
\date{\today}
\maketitle

\section{Introduction}
Collections of trapped atomic ions have been used for analog quantum simulation of equilibrium properties and non-equilibrium dynamics of spin models for nearly twenty years \cite{Monroe2021RevModPhys}. They are also a leading candidate for the development of fault-tolerant quantum computers \cite{ryan-anderson_realization_2021} and have been used to demonstrate the lowest two-qubit gate error to date \cite{clark_high-fidelity_2021}. Entangling operations remain the largest source of error, and characterizing the causes of these errors is key to performing larger and deeper quantum simulations and computations in the near term. Most prior research has focused on characterizing errors in two-qubit gates which generate maximally entangled Bell states. A common method for generating this entanglement is the M{\o}lmer-S{\o}rensen (MS) interaction \cite{Sorensen2000}, which entangles the spin state of ions through their collective normal modes of motion. Previous studies have examined two-qubit MS interaction errors due to thermal vibrational states of the ions \cite{Kirchmair2009thermal}, residual spin-motion entanglement \cite{Leung2018}, motional errors \cite{Sutherland2022motionalerrors}, laser coherence \cite{day_limits_2022}, and fast local oscillator noise \cite{Nakav2022fastnoise}. 
\par
Recently, variational approaches with classically optimized control parameters have been used to expand what can be simulated with both analog \cite{kokail_self-verifying_2019} and digital \cite{nam_ground-state_2020} quantum hardware. Replacing fixed two-qubit ``digital'' gates with continuously-variable many-qubit ``analog'' interactions expands the range of simulations and computations that can be performed with low-depth circuits \cite{arrazola_digital-analog_2016,parra-rodriguez_digital-analog_2020}. One application is the quantum approximate optimization algorithm (QAOA) \cite{farhi2014quantum}, which is often viewed as a leading candidate algorithm for near-term quantum computers \cite{Preskill2018NISQ}. QAOA has been a topic of contemporary research including experiments \cite{Pagano2020PNAS, ebadi_quantum_2022, Google2021QAOA}, theory \cite{Farhi2020SK, wurtz2021cd, hadfield2019quantum,tate2021classically,herrman2021lower}, and simulations \cite{zhou_quantum_2020, Lotshaw2021bfgs, akshay2020reachability}, with applications ranging from combinatorial optimization to quantum simulation \cite{ho2019efficient,lotshaw2022simulations,wiersema2020exploring}. In digital approaches to QAOA, it is expected that two-qubit gate error rates of $10^{-5}-10^{-6}$ will be necessary for quantum advantage \cite{GarciaPatron2021limitations,lotshaw2022scaling,Weidenfeller2022scaling,GonzalezGarcia2022propogation}, while a recent proposal for QAOA with many-qubit MS interactions uses far shallower circuits in certain contexts \cite{Rajakumar2020MSQAOA}, potentially avoiding the digital limitations.
\par
In this work, we develop a detailed simulation model for many-qubit MS interactions to understand noise that is present in analog experiments. Prior error modeling has focused largely on two-qubit gates \cite{Kirchmair2009thermal, Leung2018, Sutherland2022motionalerrors, day_limits_2022, Nakav2022fastnoise} as well as noise and control in dynamics of many trapped ions \cite{bentley2020numeric,PhysRevA.87.042101}. Here, we numerically simulate analog control of the MS interaction on up to six ions.  We show that our global MS interaction model, with parameters determined from simple experimental observations, effectively reproduces the experimental dynamics. As an application of the theory, we model QAOA experiments using an analog compilation scheme and reproduce their performance as a function of the experimental parameters. This demonstrates success of the model in describing composite circuits containing MS interactions. Ultimately, we arrive at an understanding of the performance of our trapped-ion system, validated in multiple contexts.
\par
Using the model, we predict that reducing two dominant errors---measurement error and vibrational mode frequency fluctuation---by a realistic factor of ten would improve QAOA performance by eliminating $>99\%$ of the error we model. Our work complements a previous simulation analysis of noise, described in the supplementary information to Ref. \cite{Pagano2020PNAS}, for a 12-ion instance of QAOA targeting the center-of-mass (COM) mode.  In contrast with this previous work, we construct and validate our model in the context of multiple MS gates applied in isolation, and compare the influence of each noise source to the experimental behavior.  We then extend this to QAOA instances targeting non-COM vibrational modes, with non-uniform couplings corresponding to instances of the weighted MaxCut problem with positive and negative edge weights. In total, we develop a satisfactory link between experimentally observed noise parameters, performance of the MS interaction in isolation, and algorithm performance under noisy MS operations.
\par
In \refsec{sec:bkgd}, we summarize background theory of the MS interaction. We  develop our approaches to modeling known noise sources and validate the model through comparison with experiments on two-ion chains in \refsec{sec:MS-exp}. In \refsec{sec:QAOA}, we apply our error model to Max-Cut QAOA performed with up to six ions, and the implications of this work are discussed in \refsec{sec:discussion} with future applications suggested in \refsec{sec:conclusions}.
\section{Background\label{sec:bkgd}}
\subsection{Global M{\o}lmer-S{\o}rensen interaction}
In this section we describe our model for noise in globally-entangling M{\o}lmer-S{\o}rensen (MS) interactions. In contrast to an ``MS gate,'' which is a fixed digital operation that generates a two-qubit Bell state, here we refer to ``global MS interactions'' to emphasize that we are considering many-qubit analog operations, which may produce a continuum of states depending on the control parameters. The MS interaction uses electron-phonon coupling to generate entanglement between electrons, where the phonons are ideally disentangled from the electrons at the end of a well-controlled interaction.  To treat the interaction exactly, we consider a tensor product basis of electronic and vibrational states. We use lower-case indices to refer to single-ion electronic states $\vert z_i\rangle \in \{\vert 0_i\rangle, \vert 1_i\rangle\}$ and single harmonic oscillator normal vibrational modes of the ions in the Fock basis $\vert \nu_m\rangle$, while using capital indices to denote joint electronic states $\vert z_I \rangle = \bigotimes_{i=0}^{n-1} \vert z_i\rangle$ and joint vibrational states $\vert \nu_M \rangle = \bigotimes_{m=0}^{n-1} \vert \nu_m\rangle$ of $n$ trapped ions.  
\par
We focus on trapped ion dynamics with globally-illuminating M{\o}lmer-S{\o}rensen operations that entangle all ions. 
In our setup, the MS interaction acts on the joint state of electronic and vibrational states of the ions with the propagator \cite{Kim2009Ising,Islamthesis} 
\begin{equation} \label{MS} U_\mathrm{MS}(t) =e^{ -i \sum_{i<j} \chi_{i,j}(t) X_iX_j} \prod_{m=0}^{n-1} D_m\left(\sum_{i=0}^{n-1} \alpha_{i,m}(t)X_i\right).\end{equation}
Here $i$ and $j$ indices refer to electronic basis states, $\alpha_{i,m}(t)X_i$ is a component of the displacement of mode $m$ which depends on the electronic state through $X_i$, and the spin-dependent geometric phase $\chi_{i,j}(t)X_iX_j$ is accumulated during these displacements, see Appendix \ref{ms-detail} for details. 
\par
\begin{figure}
    \centering
        \includegraphics[width=7cm,height=9cm,keepaspectratio]{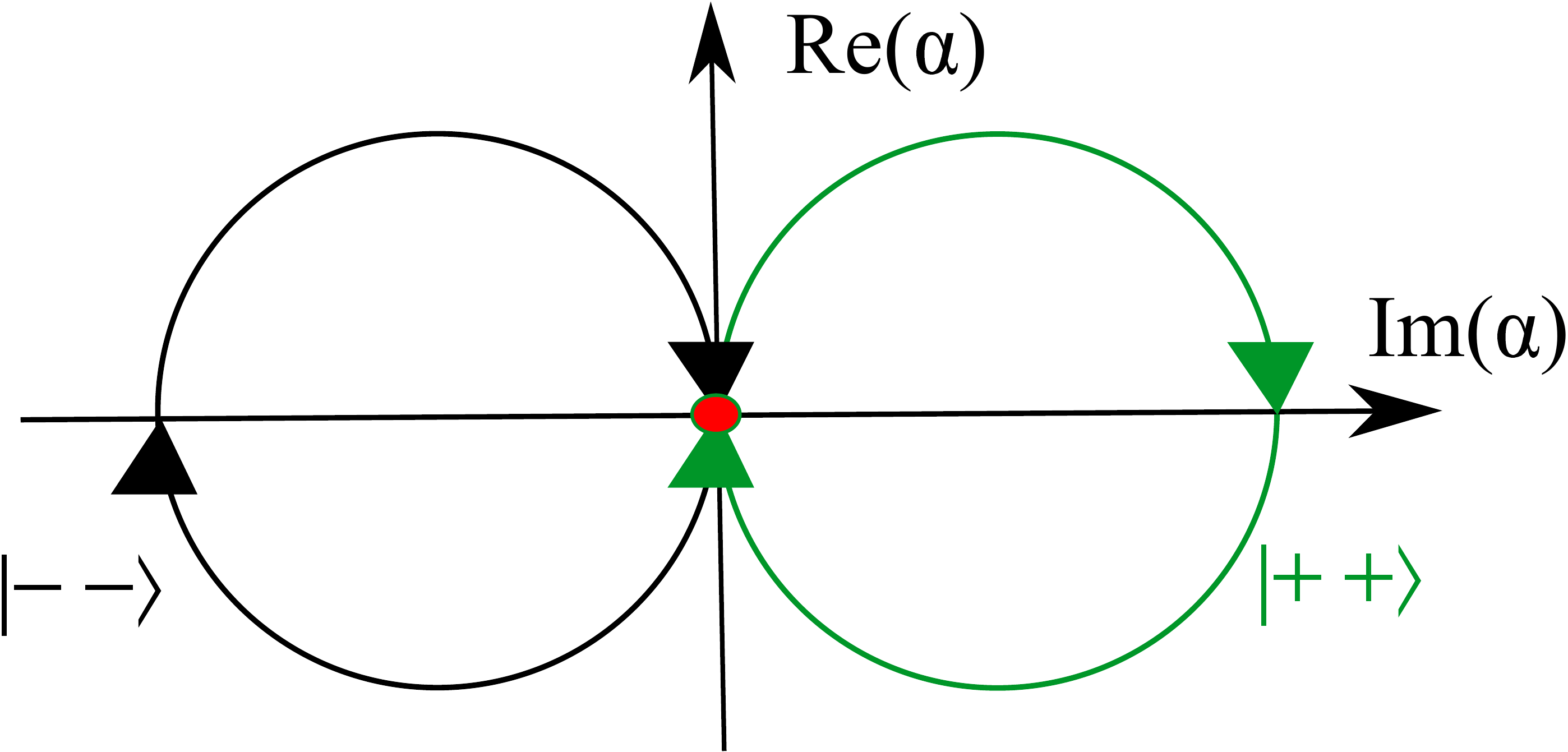}
    \caption{Schematic vibrational displacement of the COM mode for a single MS interaction applied to two ions.  The vibrational displacements trace loops in phase space that depend on the electronic states $\vert ++\rangle$ and $\vert --\rangle$.  At certain times each displacement returns to the origin, leaving an ion-ion entangling gate.}
    \label{MS schematic}
\end{figure}

The displacement operators $D_m\left(\sum_i \alpha_{i,m}(t)X_i\right)=e^{\left(\sum_i \alpha_{i,m}(t)X_i\right)a_m^\dag - \left(\sum_i \alpha_{i,m}(t)X_i\right)^*a_m}$ generate coherent states of the vibrational modes that are driven along ``loops'' in phase space through time-dependent displacements that depend on the electronic state (Fig.~\ref{MS schematic}).  
When each mode returns approximately to the origin, then $\prod_mD_m \approx \mathbb{1}$ and the MS interaction reduces to a globally-entangling interaction on the electronic levels only.  At long times (described in Appendix \ref{ms-detail}), this approximately matches evolution $e^{-i H t}$ under an Ising Hamiltonian $H = \sum_{i<j} J_{i,j}X_iX_j$ with \cite{Kim2009Ising}
\begin{equation} \label{Jij} J_{i,j} = \Omega^2 \sum_m \frac{\eta_{i,m}\eta_{j,m}}{\mu^2-\omega_m^2} \omega_m, \end{equation}
where $\Omega$ is the Raman Rabi rate (assumed constant for all ions due to uniform global illumination), $\eta_{i,m}$ are Lamb-Dicke parameters, $\mu$ is the bichromatic detuning with respect to qubit resonance, and $\omega_m$ is the frequency of vibrational mode $m$. In this work the MS interaction is detuned close to a single ``target" mode $m_t$, with detuning $|\mu-\omega_{m_t}| \ll \omega_{m_t}$. The $J_{i,j}$ coupling parameters are largely determined by the target mode, since the summands in \refeq{Jij} scale as $\omega_m/(\mu-\omega_{m})$ for each mode.
\par
The global MS interaction has been used to generate Ising evolution in previous works focused on Hamiltonian simulation \cite{Monroe2021RevModPhys} and the quantum approximate optimization algorithm \cite{Pagano2020PNAS}, where the MS interaction was detuned outside the center-of-mass mode leading to a power law decay of Ising interactions with distance between ions. In contrast, in our experiments we tune the MS detuning close to a targeted mode in the middle of the mode spectrum leading to both positive and negative $J_{i,j}$.
\par
To understand errors arising from MS interactions under realistic sources of experimental noise, we consider generic interaction times at which undesired spin-motion entanglement persists, and the motional states of the ions do not return to the origin in Fig.~\ref{MS schematic}.  We begin with an initial state $\vert \Psi_{I,M}\rangle = \vert \psi_I\rangle \vert 0_M\rangle$ that is a separable state of the ground vibrational modes $\vert 0_M\rangle$ and a generic state of the electronic levels $\vert \psi_I\rangle = \sum_{x_I} c_{x_I} \vert x_I\rangle$ expressed in the $n$-ion Pauli-$X$ eigenbasis $\{\vert x_I\rangle\} = \{\bigotimes_{i=0}^{n-1} \vert x_i\rangle\}$. The total state $\vert \Psi_{I,M}\rangle$ evolves under the MS interaction to
\begin{equation} \label{Psi MS} U_\mathrm{MS}(t) \vert \Psi_{I,M}\rangle = \sum_{x_I} c_{x_I} e^{-i \chi(t,x_I)} \vert x_I \rangle \vert \alpha_M(t,x_I)\rangle, 
\end{equation}
where $e^{-i \chi(t,x_I)}$ is the geometric phase for $\vert x_I\rangle$, with $\chi(t,x_I) = \langle x_I \vert \sum_{i,j} \chi_{i,j}(t) X_i X_j \vert x_I\rangle$, and $\vert \alpha_M(t,x_I)\rangle = \bigotimes_{m=0}^{n-1} \vert \alpha_m(t,x_I)\rangle$ is a product of vibrational mode coherent states entangled with the electronic state $\vert x_I\rangle$.  The electronic qubit states alone are described by the reduced density operator \cite{NielsenChuang}
\begin{equation} \label{ion RDM} \rho_I = \sum_{x_I,x_I'}  \vert x_I \rangle c_{x_I}c_{x_I'}^* e^{-i (\chi(t,x_I)-\chi(t,x_I'))} \epsilon_M(t,x_I,x_I')\langle x_I'\vert\end{equation}
where $\epsilon_M(t,x_I,x_I') = \prod_{m=0}^{n-1} \langle \alpha_m(t,x_I')\vert\alpha_m(t,x_I)\rangle$ quantifies decoherence from ion-vibration entanglement.  When all vibrational modes return to the origin, then $\epsilon_M(t,x_I,x_I')=1$ and $\rho_I$ is a pure electronic state, but if a vibrational mode remains excited, then $|\epsilon_M(t,x_I,x_I')| < 1$ and $\rho_I$ is a mixed reduced state due to residual ion-vibration entanglement.  
\subsection{Trapped-ion experiments}
\begin{figure}
    \centering
        \includegraphics[width=8cm,height=11cm,keepaspectratio,trim={0cm 0cm 0cm 0cm }]{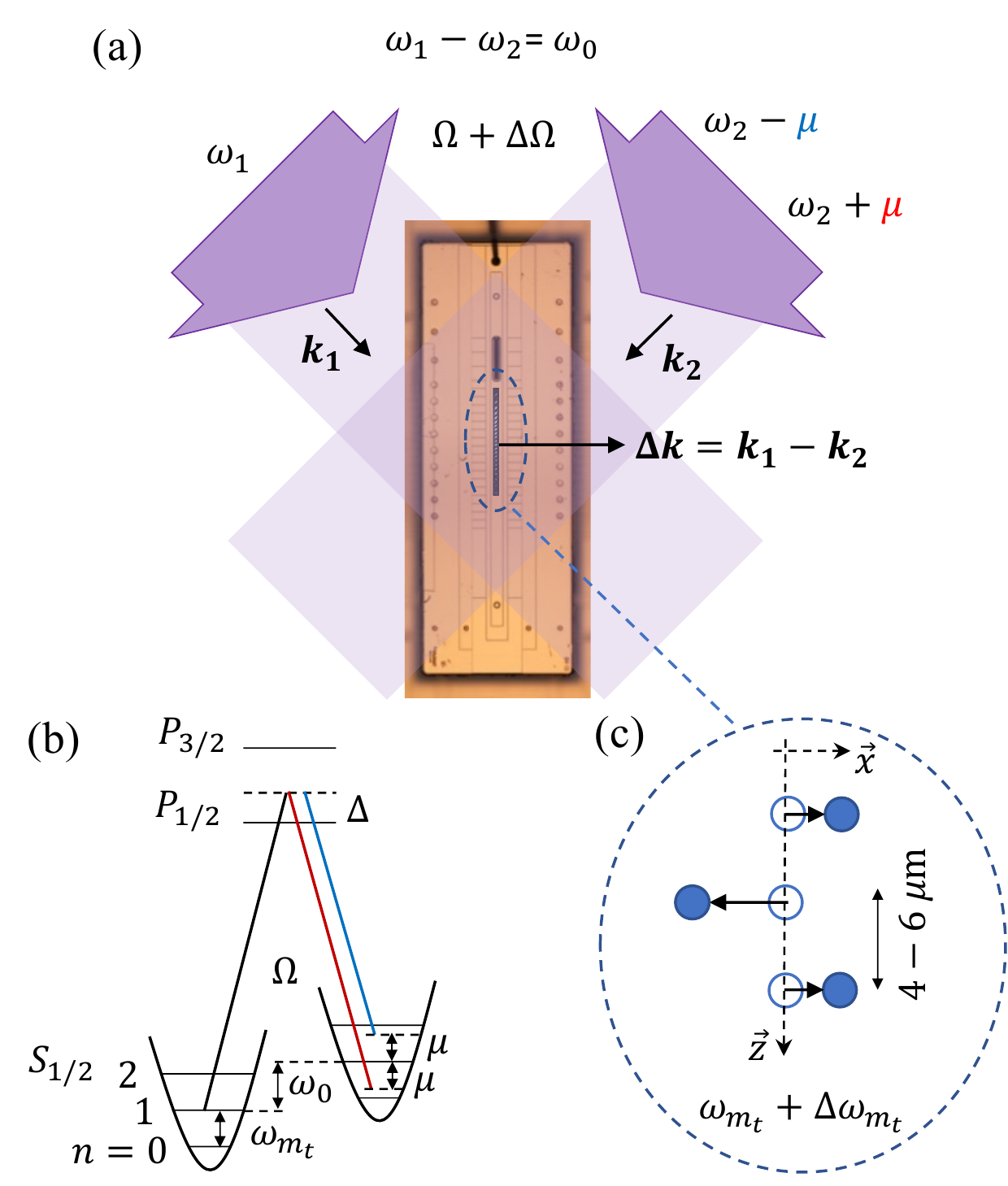}
    \caption{
    Schematic of the experimental setup for the MS interaction. (a) Top view of the ion trap~\cite{Guise2015} showcasing the wide global 355-nm Raman beams with optical frequencies $\omega_{1}, \omega_{2}+\mu$ (red sideband excitations), and $\omega_{2}-\mu$ (blue sideband), with $\mu$ the Raman detuning with respect to the spin resonance frequency $\omega_{0}$. (b) Energy level diagram of a single $^{171}\mathrm{Yb}^{+}$ ion and phonon levels of the targeted radial vibrational mode. (c) Sketch of a three-ion chain showing the ``zig-zag'' vibrational mode, as targeted in the experiment of Fig.~\ref{heatmaps 3 ion}. Individual noise sources are pictured: Fluctuations in the vibrational mode  $\Delta \omega_{m_{t}}$ in (c), while  laser power fluctuations change the Raman Rabi rate $\Delta \Omega$ in (a).
    }
    \label{expt figure}
\end{figure}

We validate our MS interaction noise model with experiments on up to six $^{171}$Yb$^+$ ions. Our implementation of the MS interaction is shown in Figure~\ref{expt figure}. The ions are trapped above a GTRI-Honeywell ball-grid array surface-electrode ion trap \cite{Guise2015}. The qubits are encoded in the hyperfine ground ``clock" states $\ket{0} \equiv {^2}S_{1/2} \ket{F=0,m_F=0}$ and $\ket{1} \equiv \ket{F=1,m_F=0}$, and state preparation and readout follow Ref.~\cite{Olmschenk2007}. The experimental apparatus is similar to that described in \cite{Herold2016}, except for one important difference: a pair of wide global 355-nm Raman beams intersecting at 90 degrees uniformly illuminates all ions and drives the MS interaction on the radial modes parallel to the trap surface (Figure~\ref{expt figure}(a)). The center-of-mass mode in the axial direction has a frequency of approximately $2\pi\times 0.5$~MHz leading to ion separations of $4-6~\mu$m. Figure~\ref{expt figure}(c) depicts the lowest energy radial mode (often called the``zig-zag'' mode as every ion moves in the opposite direction from its neighbors) of a three-ion chain. Readout is performed by imaging each ion onto a separate element of a multichannel photomultiplier tube (PMT). More details on the basic approach can be found in Ref.~\cite{Monroe2021RevModPhys}.

Further details of the observed mode frequencies, detunings, and additional experimental controls for QAOA are given in Appendix \ref{gamma-setting}. The Raman Rabi rates $\Omega$ in these experiments are inferred following the procedures of Appendix \ref{Rabi rate appendix}.

\section{Noise characterization and modeling}\label{sec:MS-exp}
\begin{figure*}[!htb]
        \includegraphics[width=\textwidth,height=20cm,keepaspectratio,trim={0cm 0cm 0cm 0cm }]{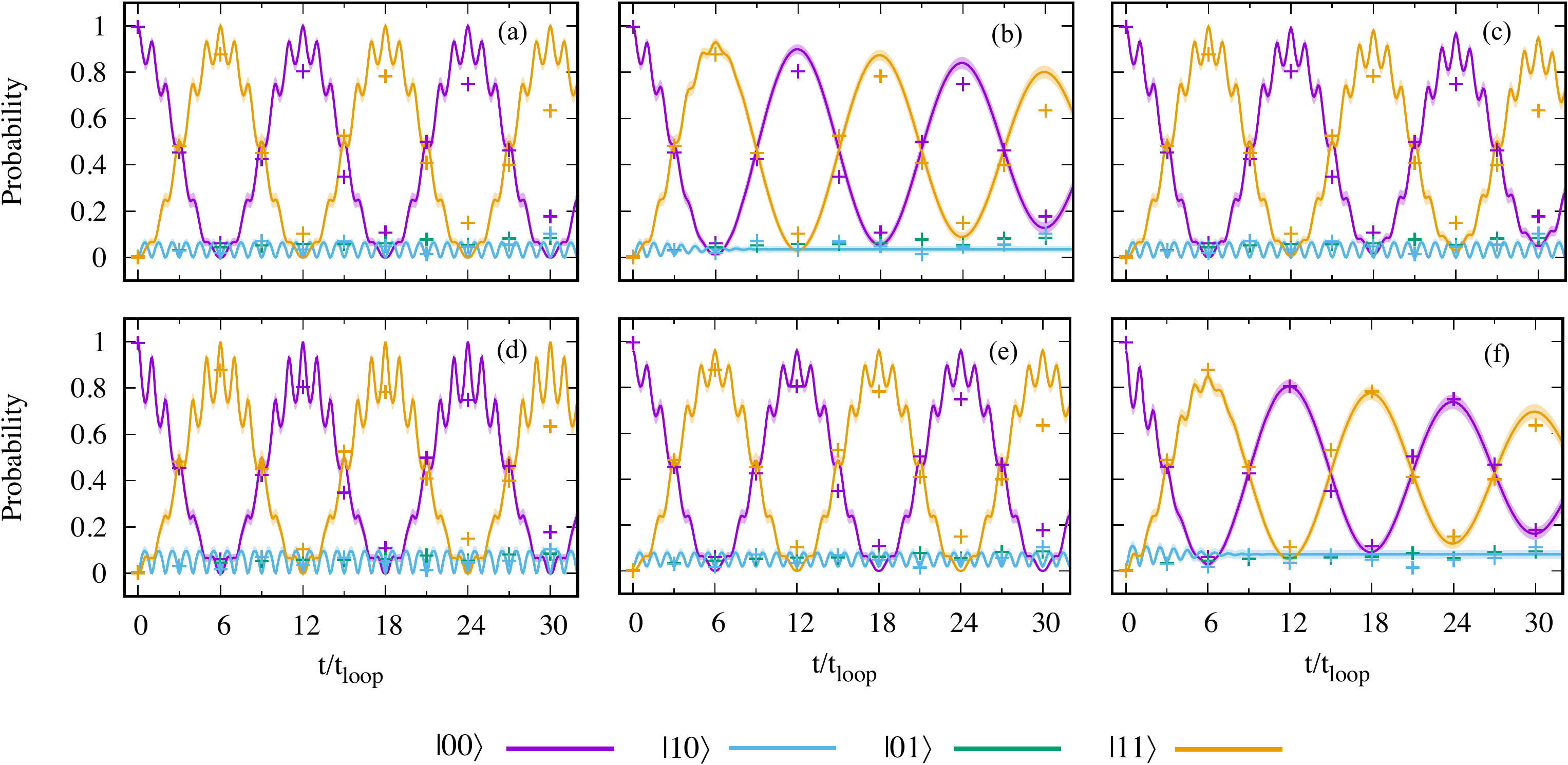}
    \caption{Modeling MS sequences on two ions. The experimental data (crosses) with $S= 200$ shots are the same in each panel. The theory (lines) show (a) the ideal dynamics. Subsequent panels show dynamics under individual noise sources: (b) fluctuations in the target mode frequency, (c) fluctuations in laser power, (d) thermal occupation of the vibrational modes, (e) SPAM error. In (f) we combine all sources of noise together; agreement with experiment is quantified by the reduced chi-squared statistic $\chi^2_\mathrm{red}=2.11$ (not to be confused with the MS coupling $\chi_{i,j}$). Note that the $\ket{01}$ trace is always identical to the $\ket{10}$ trace.}
    \label{2ion MS}
\end{figure*}

In this section we develop a noise model that accounts for experimental observations of two-ion dynamics under sequences of MS interactions. We probe experimental dynamics for two ions by applying sequences of MS gates with short $1~\mu$s time gaps between successive applications.  The experiment was calibrated to produce a Bell state after a single MS application, while multiple applications of this interaction should closely reproduce continuous time behavior. For comparison with each source of noise, the same experimental results with two ions are shown as crosses in each panel of Fig.~\ref{2ion MS}. The observed experimental populations of $\vert 00\rangle$ and $\vert 11\rangle$ decay at long times, while populations in $\vert 01\rangle$ and $\vert 10\rangle$ appear to saturate to approximately steady non-zero values.
\subsection{Ideal behavior and error bars}
When applied to the $\ket{00}$ state, the ideal continuous-time evolution of the MS interaction produces approximately sinusoidal oscillation between the $\ket{00}$ and $\ket{11}$ states as shown in the solid curves in Fig.~\ref{2ion MS}(a). Deviations from sinusoidal behavior, and populations in $\ket{01}$ and $\ket{10}$, are both due to ion-vibration coupling in the displacement operators $D_m$ which produce fast oscillations with period $t_{\mathrm{loop}}=2\pi/|\mu-\omega_{m_t}|$.  When $t/t_\mathrm{loop}$ is an integer, the targeted vibrational mode has returned to the origin in phase space, and the evolution is close to the ideal $X_0X_1$ with negligible entanglement to the motional modes.
\par
Throughout this paper we quantify the expected sampling error using the standard error of the mean, computed from the simulated density operator $\rho_I$ (Eq.~(\ref{ion RDM})) for the number of shots $S$ used in the experiments (see Appendix \ref{finite sampling} for details).  Thus, we compare the experimentally determined mean probabilities against the mean $\pm$ standard error of the simulation results, shown by colored bands in Figs.~\ref{2ion MS}$-$\ref{heatmaps 6 ion}, as an alternative to comparing mean simulation results against the experimental mean $\pm$ an estimated experimental sampling error. If the experiment is consistent with the theory, then the statistics from the experiment should be consistent with statistical expectations from the theory, and this is quantified by standard error of the mean from $\rho_I$. The decaying experimental data points (crosses) in Fig.~\ref{2ion MS}(a) are not consistent with the finite sampling errors expected from these noiseless simulations, so we consider additional noise sources below. 
\par
We model the effect of individual noise sources (plotted independently in Fig.~\ref{2ion MS}(b)-(e)), which are known from the experiment and described in the following subsections. When combined, these sources of noise yield a final composite model that matches experimental observations in Fig.~\ref{2ion MS}(f). Note  we do not fit parameters against results in  Fig.~\ref{2ion MS} to parameterize our model. Instead, we characterize noise in independent experimental measurements and use the observed measurements to define our noise model.  This direct approach allows us to clearly assess whether our understanding of experimental noise sources, as measured in independent experiments, suffices to explain the observed MS dynamics.  
\subsection{Vibrational mode frequency fluctuations\label{mode-fluctuations}}
Instability in the vibrational mode frequencies leads to variations in the displacements $\alpha_{i,m}$ and the geometric phases $\chi_{i,j}$ in the MS interaction (see Eq.~(\ref{Jij}) and Appendix~\ref{ms-detail}). Repeated measurements of the vibrational mode frequencies show that the mode frequencies are unstable at the $200$ ppm level. For context, active stabilization of the rf trap potential has produced better than 10 ppm stability \cite{johnson_active_2016}. 
\par
We model fluctuations in the target mode frequency as random Gaussian variates $\Delta \omega_{m_t}$, such that for each shot $\omega_{m_t} \to \omega_{m_t} + \Delta \omega_{m_t}$ with probability density $p(\Delta \omega_{m_t}) = (2\pi \sigma^2)^{-1/2} \exp(-\Delta \omega_{m_t}^2/2\sigma^2)$.  A similar type of model for how such errors influence $\chi_{i,j}$ has been considered in analytical bounds for precision errors in QAOA \cite{Quiroz2021precison}, though that work did not consider the vibrational-mode quantum states as we do here.  We model the expected ion reduced density operator for an ensemble of experiments each with a distinct static random fluctuation $\Delta \omega_{m_t}$ in the target mode $m_t$ as \cite{NielsenChuang}
 \begin{equation} \label{mode errors} \rho_I = \frac{1}{N}\int_{-3\sigma}^{3\sigma} d(\Delta\omega_{m_t}) p(\Delta\omega_{m_t}) \rho_I(\Delta\omega_{m_t}), \end{equation}
where $\sigma = 2\pi \times 0.3$ kHz is the standard deviation observed in repeated measurements of the target mode frequency over tens of minutes. We evaluate the integral numerically as a Riemann sum over 1000 evenly spaced intervals. The finite bounds for the integration of $[-3\sigma,3\sigma]$ include 99.73\% of the probability density of a true Gaussian on $(-\infty,\infty)$; we normalize our final state through $N$ to account for the discrepancy. Drifts in the mean vibrational frequencies also affect performance, though we do not include this in the present work.
\par
Figure \ref{2ion MS}(b) shows dynamics with fluctuations in the frequency $\omega_{m_t}$ of the target mode.  A fluctuation $\omega_{m_t} \to \omega_{m_t} + \Delta \omega_{m_t}$ changes the true loop time from the value $t_\mathrm{loop} = 2\pi/|\mu-\omega_{m_t}|$ assumed in the experiments; the rate of oscillations will increase or decrease depending on the sign of $\Delta \omega_{m_t}$.  In our model we are integrating over varying fluctuations, corresponding to an average over experimental preparations and measurements with different fluctuations.  The average over these varying rates of evolution causes a decay in the oscillations between $\vert 00\rangle$ and $\vert 11\rangle$, which becomes more pronounced at longer times, where the evolutions under different fluctuations become further out of phase.  This also causes the populations in $\vert 01\rangle$ and $\vert 10\rangle$ to approach steady values consistent with averages over their small-oscillation amplitudes. The mode frequency noise captures much of the discrepancy between ideal evolution and experiment, however, it does not yield quantitative agreement.  For this, we consider further relevant noise sources.
\subsection{Laser power fluctuations} \label{power fluc}
Fluctuations in the laser power $P_j$ of the two Raman beams that drive the MS interaction lead to fluctuations of the Rabi rate $\Omega$, which is proportional to $\sqrt{P_1 P_2}$.
We model fluctuations in $\Omega$ using a similar methodology to the treatment of errors in the mode frequencies.  We consider an ensemble of model realizations with a Gaussian probability distribution of static errors $\Delta\Omega$ for each shot and numerically integrate the reduced density operator analogous to Eq.~(\ref{mode errors}) to determine the expected influence of these errors. We estimate a standard deviation of $1.5\%$, based on measurements of the beam power fluctuations near the ion trap. Here we assume the beam powers fluctuate together, which neglects AC Stark shifts that may occur when the beam powers and associated Rabi rates fluctuate separately.  Using the model of Ref.~\cite{Morong2022dynamical}, we estimate that Stark shifts from separately fluctuating beams have a negligible effect in Fig.~\ref{2ion MS}, though a detailed characterization of the influences of these errors for increasing numbers of ions and timescales is an open topic for future work.
\par
Figure \ref{2ion MS}(c) isolates the influence of variations in the Rabi rate $\Omega$. The errors lead to gradual damping of the oscillations between $\vert 00\rangle$ and $\vert 11\rangle$, observable at longer times, while the populations in $\vert 01\rangle$ and $\vert 10\rangle$ are mostly unaffected.  
\subsection{Thermal initial states of the vibrations} \label{thermal}
Imperfect sideband cooling leads to a non-zero average occupation of the vibrational modes. As a simple model for such excitations we consider an initial thermal state for each vibrational mode $m$ \cite{Roos2008amplitude},
\begin{equation} \rho_m(\overline{\nu}_m) = \frac{1}{\overline{\nu}_m+1} \sum_{\nu_m=0}^\infty \left(\frac{\overline{\nu}_m}{\overline{\nu}_m+1}\right)^{\nu_m} \vert \nu_m \rangle\langle \nu_m \vert, \end{equation} 
where $\overline{\nu}_m$ is the expected mode occupation number. We consider an initial total state of all vibrational modes as $\rho_M = \bigotimes_{m=0}^{n-1} \rho_m(\overline{\nu}_m)$.  Theoretically the $\overline{\nu}_m$ could vary for different modes depending on how effectively they are cooled in experiments, but for simplicity here we consider identical occupation numbers $\overline{v_m}=\overline{v}$ for each mode. In each experiment, all radial modes are cooled to the ground state through sideband cooling with average occupation $\overline{v}<0.5$ measured by observing sideband ratios. While the observed heating rate for most modes is negligible, this model does not incorporate the known heating of the center-of-mass mode; however since we are much further detuned from this mode, heating is not expected to be a significant source of error for the experiments presented here.
\par
The thermal vibrational states influence the ion-vibration dynamics and hence the ion reduced density operator in Eq.~(\ref{ion RDM}). This modifies the decoherence term to \cite{Roos2008amplitude}
\begin{align} \epsilon_M^{(\overline{\nu}_m)}(t,x_I,x_I') = \prod_m & \exp(i\mathrm{Im}[\alpha_m(t,x_I)\alpha_m^*(t,x_I')]) \nonumber \\
& \times |\langle \alpha_m(t,x_I')\vert\alpha_m(t,x_I)\rangle|^{2\overline{\nu}_m+1} \end{align} 
Thus, decoherence between states $\vert x_I \rangle$ and $\vert x_I'\rangle$ increases exponentially with $\overline{\nu}_m$.  This decoherence is expected to significantly degrade performance when either $\alpha_m(t,x_I)$ or $\overline{\nu}_m$ is large as the MS interaction becomes more sensitive to calibration errors and parameter fluctuations.
\par
Figure~\ref{2ion MS}(d) considers a thermal initial state for each vibrational mode with mean occupation numbers $\overline{\nu}=0.5$.  The more rapidly oscillating populations in $\vert 01 \rangle$ and $\vert 10\rangle$ are increased relative to Fig.~\ref{2ion MS}(a), while the populations in $\vert 00\rangle$ and $\vert 11 \rangle$ have correspondingly larger superimposed rapid oscillations away from the ideal two-level-system limit of sinusoidal oscillations.  The vibrations still become effectively disentangled from the ions at integer $t/t_\mathrm{loop}$, where $\epsilon_{M} \approx 1$, but the differences in amplitude are significant at other times. Such non-ideal times may be present experimentally, for example, due to mode frequency fluctuations that cause deviations in the loop times away from the experimentally assumed values.
\subsection{SPAM error\label{sec:SPAM}}
State preparation and measurement (SPAM) errors are primarily due to electronic crosstalk between neighboring channels of the 32-channel PMT used for readout.  The crosstalk in our multi-qubit detection could be improved by using a fiber array coupled to separate PMTs \cite{clark_engineering_2021} or by using a camera. Beyond the errors considered here, detection of the ${}^{171}$Yb$^+$ qubit is fundamentally limited to about $10^{-3}$ infidelity by off-resonant transitions between the qubit states during readout \cite{Noek2013}.
\par
To account for SPAM errors in our model, we use SPAM matrices $M$ that were measured experimentally as described in Appendix \ref{SPAM appendix}. The matrix elements  $M_{z_I',z_I}$ are the conditional error probabilities $P(z_I' | z_I)$ to observe $\ket{z_I'}$ when preparing $\ket{z_I}$. This modifies the theoretical measurement probabilities we compute from the density operator $\rho_I$, expressed as a vector $\vec P$ with elements $P(z_I) = \langle z_I \vert \rho_I \vert z_I\rangle$, to give noisy measurement results $\vec P_\mathrm{SPAM} = M \times \vec P$. The observed SPAM errors can be approximated by 2\% independent bit-flip errors, and we use this bit-flip model for our two-ion results as we did not measure the SPAM matrix in this case.  
\par
Figure~\ref{2ion MS}(e) shows the influence of SPAM errors on the dynamics.  These reduce the contrast of the populations, decreasing large populations and increasing smaller populations. The net effect is relatively small for the two-ion case, though it becomes increasingly significant for greater numbers of ions.
\subsection{Total composite noise model}
Figure \ref{2ion MS}(f) shows the influence of all sources of noise together.  The dynamics resemble the dominant source of noise from mode fluctuations in Fig.~\ref{2ion MS}(b), while the additional sources of noise give a slightly more rapid decay to the $\vert 00\rangle$ and $\vert 11\rangle$ populations along with a slightly larger steady value for the populations in $\vert 01\rangle$ and $\vert 10\rangle$. To quantify agreement we consider the reduced chi-squared statistic
\begin{equation} \label{chisquared} \chi^2_{\mathrm{red}} = \frac{1}{A} \sum_{a=1}^A \frac{(P_a^\mathrm{sim}-P_a^\mathrm{exp})^2}{\Delta P_a^2}\end{equation}
where $P_a^\mathrm{expt}$ and $P_a^\mathrm{sim}$ are the experimental and simulated measurement probabilities, respectively, $\Delta P_a^2$ is the squared standard error of the mean, and $A$ is the total number of observations (in general $\chi^2_\mathrm{red}$ also depends on the number of fit parameters in a given model, but we do not use any fit parameters in this work).  The $\chi^2_{\mathrm{red}}$ measures the average squared deviation between theory and experiment relative to the expected variance at each point.  This should be close to one for a model that describes the data accurately without overfitting.  For the composite model of Fig.~\ref{2ion MS}(f), we find reasonable quantitative agreement with $\chi^2_{\mathrm{red}}=2.11$.
\section{Quantum Approximate Optimization Experiments and Modeling\label{sec:QAOA}}
Having validated our model for the MS interaction in \refsec{sec:MS-exp}, we now apply our error model to QAOA with three and six ions. 
\subsection{Quantum Approximate Optimization Algorithm}
We consider the quantum approximate optimization algorithm (QAOA) \cite{farhi2014quantum} as a use case for global MS operations and to assess our noise modeling. QAOA is an approach to find approximate solutions to NP-hard combinatorial optimization problems. These problems are defined by a classical cost function $\mathcal{C}(z_I)$ \cite{lucas2014mapping}, with $z_I = (z_0,\ldots, z_{n-1})$ an $n$-bit string with $z_i \in \{1,-1\}$.  The problem instance is encoded into an operator $C$ with an eigenspectrum that contains the set of classical cost function values $C\vert z_I\rangle = \mathcal{C}(z_I) \vert z_I \rangle$. We focus on the weighted MaxCut problem with 
\begin{equation} \label{C} C = \frac{1}{2}\sum_{i<j} w_{i,j}(\mathbb{1}-Z_i Z_j) \end{equation} 
where $w_{i,j}$ are instance-specific weights and $Z_i$ are Pauli operators.  The goal is to identify a solution $\vert z_I^*\rangle$ with the highest cost value possible.  

In QAOA one prepares a quantum state $\vert \bm \gamma,\bm\beta\rangle$ using $p$ layers of unitary operators that each alternate between Hamiltonian evolution under $C$ and under a ``mixing" Hamiltonian $B = \sum_i X_i$,
\begin{equation} \label{QAOA} \vert \bm \gamma,\bm\beta\rangle = \prod_{l=1}^p e^{-i \beta_l B} e^{-i \gamma_l C} \vert +\rangle^{\otimes n}.\end{equation} 
The $\bm \gamma = (\gamma_1,...,\gamma_p)$ and $\bm \beta = (\beta_1,...,\beta_p)$ are variational parameters chosen to maximize the objective $\langle C \rangle = \langle \bm \gamma, \bm \beta \vert C \vert \bm \gamma,\bm\beta\rangle$, such that repeated measurements return solutions $\vert z_I\rangle$ with large expected cost. For benchmarking, performance is typically quantified by the approximation ratio 
\begin{equation} \label{approximation ratio} r = \frac{\langle C \rangle-C_\mathrm{min}}{C_\mathrm{max}-C_\mathrm{min}}\,,\end{equation} 
where $0 \leq r \leq 1$ and $C_\mathrm{max}$ and $C_\mathrm{min}$ are the largest and smallest eigenvalues of $C$.
\par
We design the MS evolution to match the QAOA cost Hamiltonian evolution $\exp(-i \gamma C)$, working in the Ising limit (\refeq{Jij}) of the MS interaction (\refeq{MS}). To design the correspondence we drop the term $\mathbb{1}$ in Eq.~(\ref{C}) as it imparts a physically-irrelevant global phase. We consider MaxCut instances with dimensionless edge weights
\begin{equation} w_{i,j} = \frac{J_{i,j}}{J_\mathrm{max}},
\end{equation}
where $J_\mathrm{max} = \mathrm{max}_{i,j} |J_{i,j}|$ is the Ising coupling of largest magnitude. The QAOA $\gamma$ angle is then given by equating $\gamma (-w_{i,j}/2) = J_{i,j}t$, or
\begin{equation} \label{gamma simple} \gamma = -2J_\mathrm{max} t. \end{equation}
Finally, note the MS evolution we consider is in terms of $X_iX_j$ interactions.  To implement $Z_iZ_j$ interactions for QAOA, the MS evolution is surrounded by global $Y$-rotations as $R_Y(-\pi/2) U_\mathrm{MS}(t) R_Y(\pi/2)$ with $R_Y(\theta) = \prod_{i=0}^{n-1} e^{-i (\theta/2)Y_i}$. Following all of these steps, we have $\exp(-i \gamma C) = R_Y(-\pi/2) U_\mathrm{MS}(t) R_Y(\pi/2)$. 
\par
As a technical note, it is necessary to implement the MS interaction in such a way that the target mode oscillates through an integer number of loops in phase space, to disentangle the vibrations from the electronic states and yield the Ising interaction of \refeq{Jij}.  This is accomplished by setting the time $t$ to an integer number of loops and varying $\Omega$ to obtain a continuous range of $\gamma$ parameters; see Appendix \ref{gamma-setting} for more details. 
\par
Edge weights for the QAOA instances we consider are depicted in Fig.~\ref{Cost instances} of Appendix \ref{gamma-setting}.
\subsection{QAOA results}
\begin{figure*}
 \centering
     \includegraphics[width=\textwidth,height=\textheight,keepaspectratio]{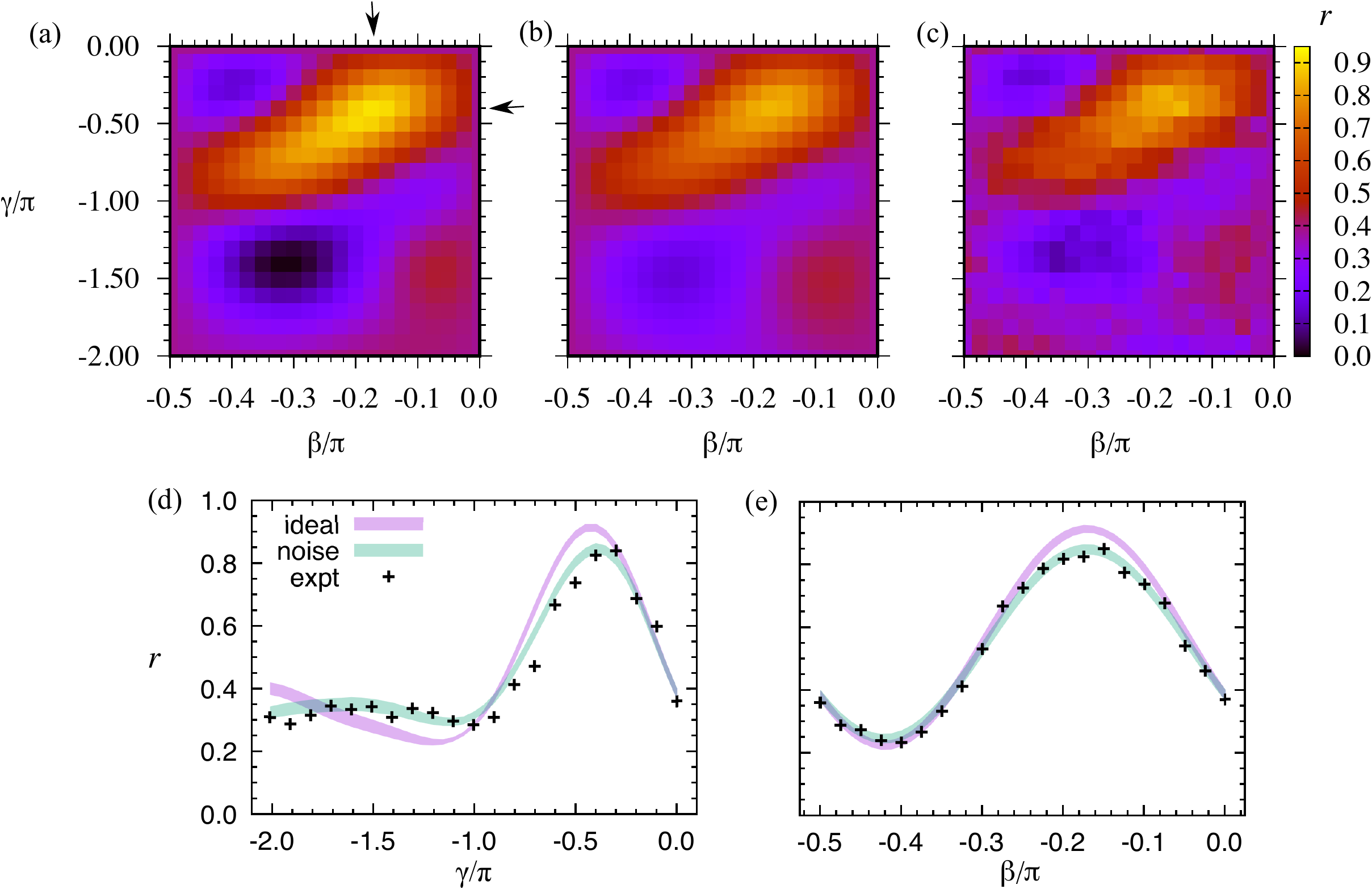}
     \caption{QAOA parameter heatmaps with three ions and $S=400$ shots per pixel in (a) the ideal case, (b) with all noise sources, and (c) the experiment. Arrows in (a) indicate the theoretical optimal parameters of $\gamma^*$ and $\beta^*$. To directly compare the noise model with the experimental data, $\gamma$ and $\beta$ slices through the heatmaps are shown for at (d) $\beta^*$ and (e) $\gamma^*$. For the curves in (d) $\chi_\mathrm{red}^2 =4.91$, for (e) $\chi_\mathrm{red}^2 =1.08$.}
     \label{heatmaps 3 ion}
\end{figure*}

We first test our model with a QAOA experiment on three ions, where the MS interaction is detuned by $-5.26$~kHz from the zig-zag mode. This gives an interaction graph with edge weights $\{1,-0.470, 1\}$. Figure \ref{heatmaps 3 ion}(a) shows the ideal approximation ratio heatmap, obtained from a simulation of QAOA dynamics in Eq.~(\ref{QAOA}), where for all simulation results we take $\langle C \rangle = \mathrm{Tr}(\rho_I C)$ in \refeq{approximation ratio}.  The approximation ratio has a maximum at angles $\gamma^*$ and $\beta^*$ located at the pixel indicated by arrows on the sides of the diagram.  Figure \ref{heatmaps 3 ion}(b) shows the heatmap generated from our simulation model with all noise sources from Section \ref{sec:MS-exp}, while Fig.~\ref{heatmaps 3 ion}(c) shows the experimentally measured heatmap.  Qualitatively, the experiment and noise model are ``damped" in comparison with the ideal case, with lower maxima and higher minima throughout the heatmaps.
\par
For a closer examination of the MS performance, we plot slices through each of these heatmaps along the pixels containing the optimal parameters $\beta^*$ and $\gamma^*$ in Fig. \ref{heatmaps 3 ion}(d) and (e) respectively.  In each figure, the semi-transparent colored bands show the expected $r$ $\pm$ the standard error of the mean as described in Appendix \ref{finite sampling}. This quantifies the expected deviation from finite sampling in the ideal and noisy cases, similar to our treatment of sampling errors in Fig.~\ref{2ion MS}. The experimental results differ significantly from the ideal case while the noise model gives a much better account of the experimental results, consistent with the expected influence of MS noise from the characterization experiments of \refsec{sec:MS-exp}.  
\par
\begin{figure*}
 \centering
     \includegraphics[width=\textwidth,height=\textheight,keepaspectratio]{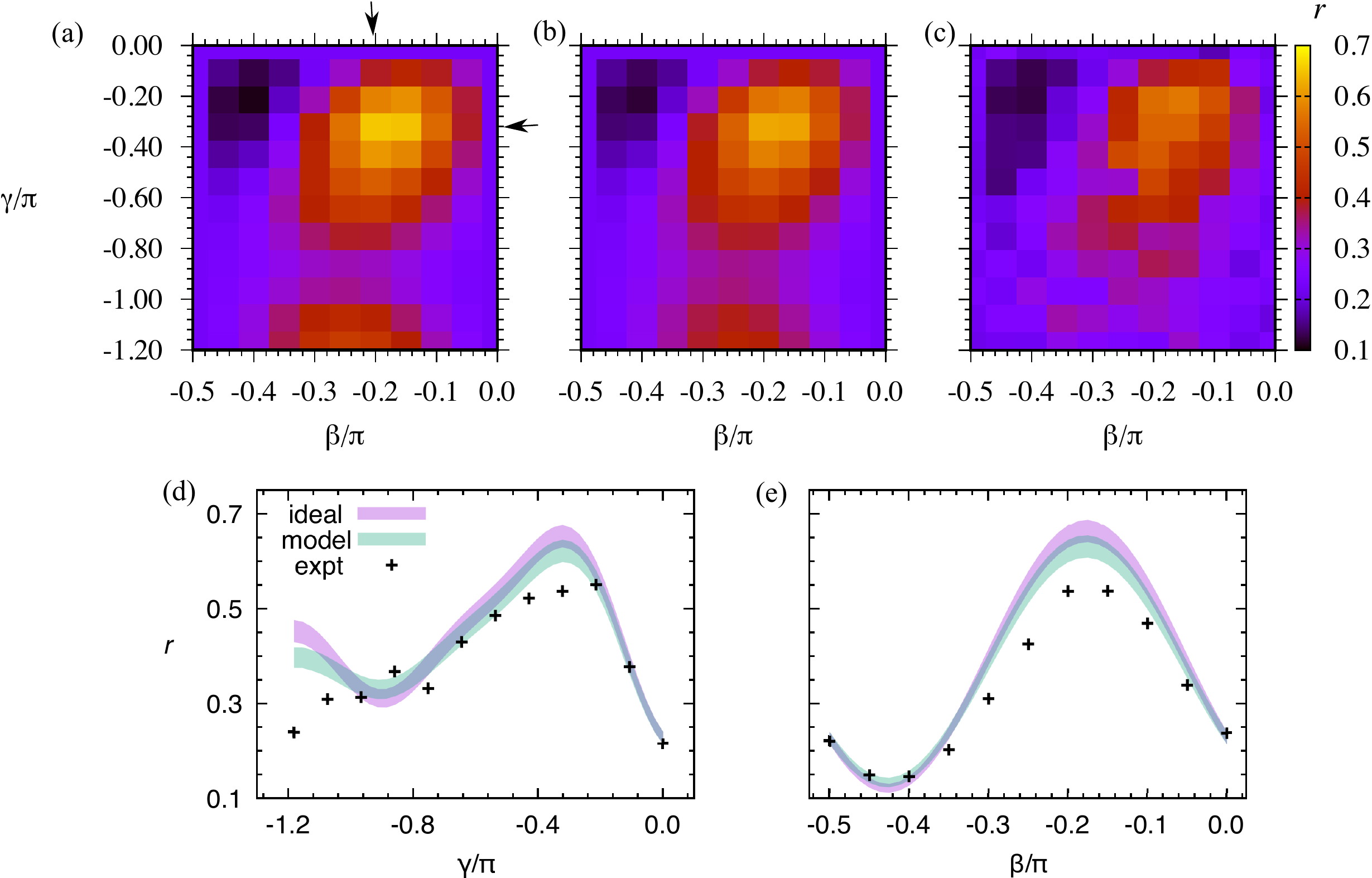}
     \caption{QAOA heatmaps with six ions and $S=256$ shots per pixel, similar to Fig.~\ref{heatmaps 3 ion}. For the curves in (d) $\chi_\mathrm{red}^2 =7.57$, for (e) $\chi_\mathrm{red}^2 =7.72$.}
     \label{heatmaps 6 ion}
\end{figure*}

Next we compare ideal, noisy simulation, and experimental heatmaps for QAOA with six ions in Fig.~\ref{heatmaps 6 ion}. Here, the MS interaction is detuned $-6.20$~kHz from the $m_t=3$ transverse radial mode, resulting in a complete interaction graph with both positive and negative edge weights; see Appendix~\ref{gamma-setting} for observed mode frequencies and calculated graph edge weights. As for the 3-ion case, the noise model captures damping in the landscape that is observed in the experiment, and captures some of the decay in maximum performance. Unexplained deviations are present in Fig.~\ref{heatmaps 6 ion}(d) and (e).

We examine the influences of individual noise sources on our results in Table \ref{performance table}.  Adding sources of noise one at a time, in order of increasing importance for the 6-ion case, we compute the approximation ratio $r^*$ for the pixel containing optimal QAOA parameters as well as the root-mean-square error (RMSE) and $\chi^2_\mathrm{red}$ between theory and experiment across all pixels in the heatmaps. Note for $\chi^2_\mathrm{red}$ this takes the expectation value $\langle C \rangle$ and squared standard error of the mean $\Delta C^2$ in place of the $P$ and $\Delta P^2$ in Eq.~(\ref{chisquared}), as these are the measured quantities relevant to the heatmap; the RMSE is computed similarly but without $\Delta C^2$.  The most important sources of noise are SPAM errors and vibrational frequency fluctuations, which give the largest reductions to RMSE, $\chi^2_\mathrm{red}$, and $r^*$, while the other error sources have minor impacts. 
\begin{table*}
\caption{\label{performance table}Comparison between experiments and noisy simulations. The first line shows the noiseless simulation, while each subsequent line includes the accumulated effect of additional noise sources.  Here RMSE is the root-mean-square error over all pixels in the simulated and experimental approximation ratio heatmaps, $\chi^2_\mathrm{red}$ is the reduced chi-squared across all pixels in the heatmaps, and $r^*$ is the approximation ratio in the pixel that contains the theoretical optimal parameters, indicated by arrows in Figs.~\ref{heatmaps 3 ion}-\ref{heatmaps 6 ion}.}
\begin{ruledtabular}
\begin{tabular}{cccccccc} 
    &\multicolumn{3}{c}{3 ion} &\multicolumn{3}{c}{6 ion} \\
    \cline{2-4} \cline{5-7}
   & RMSE & $\chi^2_\mathrm{red}$ & $r^*$ & RMSE & $\chi^2_\mathrm{red}$ & $r^*$\\
  \hline
  \multicolumn{1}{c}{noiseless} & $7.41\times 10^{-2}$ & 33.42 & $0.91\pm0.01$ & $5.61\times 10^{-2}$ & 8.24 & $0.65 \pm 0.02$\\
  \multicolumn{1}{c}{+ SPAM} & $6.61\times 10^{-2}$ & 16.57 & $0.88\pm0.01$ & $5.02\times 10^{-2}$ & 6.64 &$0.63 \pm 0.02$ \\
  \multicolumn{1}{c}{+ vib. freq. fluc.} & $5.36\times 10^{-2}$ & 8.64 & $0.86\pm0.02$ & $4.61\times 10^{-2}$ & 5.77 & $0.62 \pm 0.02$  \\
  \multicolumn{1}{c}{+ therm. init. state} & $5.26\times 10^{-2}$ & 8.06 &  $0.85 \pm 0.02$ & $4.44\times 10^{-2}$ & 5.44 & $0.62 \pm 0.02$ \\
  \multicolumn{1}{c}{+ laser power fluc.} & $5.25\times 10^{-2}$ & 7.99 & $0.85\pm0.02$ & $4.41\times10^{-2}$ & 5.37 & $0.62 \pm 0.02$ \\
  \multicolumn{1}{c}{expt.} &  & & 0.83 &  & & 0.54 \\
\end{tabular}
\end{ruledtabular}
\end{table*}

\section{Discussion\label{sec:discussion}}
Our analysis in Table \ref{performance table} indicates that SPAM and vibrational frequency fluctuations are the dominant sources of error, while the effects of thermal mode occupation and laser power fluctuations are negligible in our experiments. For our model we primarily used errors based on experimentally observed SPAM matrices, but note these can be approximated by 2\% bit-flip errors on each qubit. As the number of ions increases these errors become increasingly important, as any fixed error-per-qubit will ultimately overwhelm the results at large enough sizes. Fluctuations in the vibrational mode frequencies of the trapped ions are the dominant source of error at large interaction times. As noted in Section \ref{sec:MS-exp}, both these errors can be reduced through previously published experimental upgrades. 
\par
Using our model, we can forecast how minimizing experimental noise may improve results.  For this, we consider QAOA experiments in which the dominant noise sources are reduced by a factor of 10.  Specifically, we use vibrational frequency fluctuations of $2\pi \times 30$ Hz and a simplified model of SPAM errors treated as independent bit flips with probabilities $0.2\%$ per bit. With the factor of 10 reductions in these errors, we computed approximation ratios in our optimal pixels and find $r^*_\mathrm{noise}/r^*_\mathrm{ideal} = 0.994$ for 3 ions and $r^*_\mathrm{noise}/r^*_\mathrm{ideal}=0.992$ for 6 ions, where $r^*_\mathrm{noise}$ is the approximation ratio with all sources of error and $r^*_\mathrm{ideal}$ is the ideal noiseless approximation ratio. These compare favorably to the values computed from our analysis in Sec.~\ref{sec:QAOA}, $r^*_\mathrm{noise}/r^*_\mathrm{ideal}=0.93^{+0.03}_{-0.02}$ and $r^*_\mathrm{noise}/r^*_\mathrm{ideal}=0.95^{+0.04}_{-0.03}$ respectively.
\par
Decreasing SPAM and vibrational frequency errors will significantly reduce the experimental noise according to our model. This will likely reveal new sources of errors for future modeling and improvement. In particular, we expect characterizing drift in calibration parameters at long times to be important for reliably running QAOA for hours; slow drift can be mitigated by recalibrating periodically, however, knowing we can do so infrequently provides more time for running computations. In addition, there may be temporal correlations to the noise at short times which may require more sophisticated spectral characterization and mitigation techniques 
\cite{mundada_experimental_2022}. Fast parameter variation is not captured by our noise model in which a random parameter offset is chosen and held static for each realization. In addition, we have not considered independent beam power fluctuations, which may lead to significant AC Stark shifts in certain experimental conditions, as mentioned in Sec.~\ref{power fluc}. Using the model of  Ref.~\cite{Morong2022dynamical} we estimate that AC Stark shift contribution is negligible near the optimal parameters $(\gamma^*, \beta^*)$ in our results, though it would be useful in future work to build on the model of Ref.~\cite{Morong2022dynamical} to assess the influence of the AC Stark shift at higher orders in the Magnus expansion of the time-ordered integral that defines the propagator, including how these errors influence the vibrational mode dynamics.
\par
There are practical limitations of our modeling approach as well. Our method is based on an exact calculation of the quantum state and this scales exponentially in compute time and memory with the number of ions.  We also considered a single MS interaction, while extensions to multiple MS interactions interleaved with other operations are necessary for modeling generic quantum circuits and QAOA instances \cite{Rajakumar2020MSQAOA}. The main bottleneck in applying our approach to deeper circuits is that tracking the evolution of the vibrational modes becomes more involved.  We plan future work to address this, but an exact approach along the lines of the present work will only be feasible for sufficiently shallow circuits.
\par
An alternative to the current approach is to implement quantum channel operators, e.g.~Ref.~\cite{debroy_logical_2020}, which can be applied for arbitrary depths at the cost of approximating the influence of the vibrations or other noise sources.   Any quantum state simulation will fail for large enough numbers of qubits and depths, though analytic bounds may help guide expectations in such cases \cite{Quiroz2021precison,Coles2020nibp,GonzalezGarcia2022propogation,GarciaPatron2021limitations}. However, to perform high fidelity quantum computations at large sizes and depths it is necessary to first understand and improve noise at small sizes, and we expect the model we have developed will be useful in this context.
\section{Conclusions and Outlook\label{sec:conclusions}}
Analog many-qubit M{\o}lmer-S{\o}rensen interactions offer a promising route for implementing near-term quantum algorithms at smaller circuit depths.  We modeled these interactions under multiple sources of noise that were measured from calibration experiments, leading to a composite noise model that successfully accounted for the observed dynamics of two ions under a series of MS interactions. A defining feature of our model is that the noise parameters we use are easily identified or estimated from experiments, without fitting.  Thus we obtain a direct link between experimental observations and expected noisy performance.
\par
As an application of our approach, we modeled trapped-ion implementations of QAOA with up to six ions. Here the MS interaction is expected to be the leading source of noise and we did not include other noise sources.  Our model succeeded in accounting for decays observed in experimental QAOA performance across varying parameters, however, we also observed systematic offsets not captured by the model. These may be due to drifts in the experimental parameters, which are expected under the long operation times used in the experiments.  Noise characterization over the relevant time scales could lead to time-dependent noise models that could be readily incorporated into the current simulations to further improve their physical realism and agreement with experiment. Our modeling approach could also be applied to study the influences of noisy MS interactions in other contexts, such as Hamiltonian simulation. Extending the current approach to more complicated instances of many-qubit MS interactions is an exciting future direction.

\begin{acknowledgements}
The authors thank Joseph Wang and Brian Sawyer for helpful discussions of trapped ion physics. This material is based upon work supported by the Defense Advanced Research Projects Agency (DARPA) under Contract No. HR001120C0046.
\end{acknowledgements}

\appendix 
\section{Details of the MS interaction\label{ms-detail}}
Here we provide additional details regarding the MS interaction.  In \refeq{MS}, the geometric phase is given by \cite{Kim2009Ising}
\begin{align} \label{chi full} \chi_{i,j}(t) = & -\Omega_i\Omega_j \sum_m \frac{\eta_{i,m}\eta_{j,m}}{\mu^2-\omega_m^2}  \bigg[\frac{\mu \sin[(\mu-\omega_m)t]}{\mu-\omega_m} \nonumber\\
& - \frac{\mu\sin[(\mu+\omega_m)t]}{\mu+\omega_m} + \frac{\omega_m\sin(2\mu t)}{2\mu} - \omega_m t\bigg] \end{align}
This reduces to the Ising interaction \refeq{Jij} in the limit $t \gg (\mu/\omega_{m_t})/|\mu-\omega_{m_t}|$, as each of the first three terms in parenthesis is bounded while the final term increases $\sim t$. The phase space displacements of the vibrational mode coherent states are \cite{Kim2009Ising}
\begin{equation} \label{alpha full} \alpha_{i,m}(t) = \frac{-i \eta_{i,m} \Omega_i}{\mu^2-\omega_m^2} \left[\mu - e^{i\omega_mt}(\mu \cos(\mu t) - i \omega_m \sin(\mu t))\right].\end{equation}
These can be approximated with expressions in Ref.~\cite{Islamthesis}, which show that the time for mode $m$ to complete a phase space loop that approximately returns to the origin (Fig. \ref{MS schematic}) is $t_\mathrm{loop} = 2\pi/|\mu-\omega_{m_t}|$. With the expressions \refeq{chi full} and \refeq{alpha full} it is straightforward to compute the time-dependent electronic+vibrational pure state in \refeq{Psi MS}, and the resulting electronic reduced density operator in \refeq{ion RDM}.  The overlaps of the vibrational components entangled with different electronic states $\vert x_I\rangle$ and $\vert x_I'\rangle$, which define the $\epsilon_M(t,x_I,x_I')$ in \refeq{ion RDM}, are given analytically by
\begin{align} \label{epsilon m} \langle \alpha_m(t,x_I')\vert \alpha_m(t,x_I)\rangle & = \exp(i\mathrm{Im}[\alpha_m(t,x_I)\alpha_m^*(t,x_I')]) \nonumber\\
& \times\exp(-|\alpha_m(t,x_I)-\alpha_m(t,x_I')|^2/2).\end{align}
\section{QAOA with the MS interaction\label{gamma-setting}}
We use a microwave horn to deliver resonant microwaves for global one-qubit rotations such as in the QAOA operator $e^{-i\beta_l B}$ of \refsec{sec:QAOA}. Robust phase estimation experiments show that these operations are uniform across a six-ion chain to within $0.2\%$. With a fixed power calibrated for a $\pi/2$ gate in time $t_{\mu}$, the parameter $\beta$ is set by driving with microwaves for time $4 \beta t_{\mu}/\pi$.
\begin{table*}
\caption{\label{expt-freqs} Measured vibrational mode frequencies and MS interaction detuning from targeted mode.}
\begin{ruledtabular}
\begin{tabular}{rlll} 
  Parameter & 2-ion MS & 3-ion QAOA & 6-ion QAOA \\
  \hline
  Measured mode frequencies \\
  $\omega_0/2\pi$ & 1.7331 MHz  & 1.7328 MHz & 1.7398 MHz\\
  $\omega_1/2\pi$ & 1.6641 & 1.6635 & 1.6989 \\
  $\omega_2/2\pi$ & & 1.5615 & 1.6363 \\
  $\omega_3/2\pi$ & & & 1.5555 \\
  $\omega_4/2\pi$ & & & 1.4554 \\
  $\omega_5/2\pi$ & & & 1.3324 \\
  Targeted mode $\omega_{m_t}$ & $\omega_1$ & $\omega_2$ & $\omega_3$ \\
  MS detuning $(\mu - \omega_{m_t})/2\pi$ & -6.57 kHz & -5.26 kHz & -6.20 kHz \\
\end{tabular}
\end{ruledtabular}
\end{table*}
\par
\begin{figure}
 \centering
     \includegraphics[width=4cm,height=4cm,keepaspectratio]{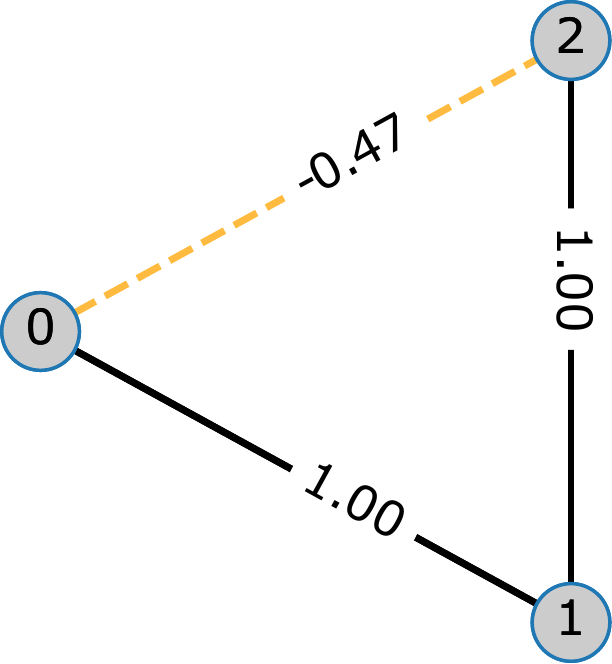}
    \includegraphics[width=6cm,height=6cm,keepaspectratio]{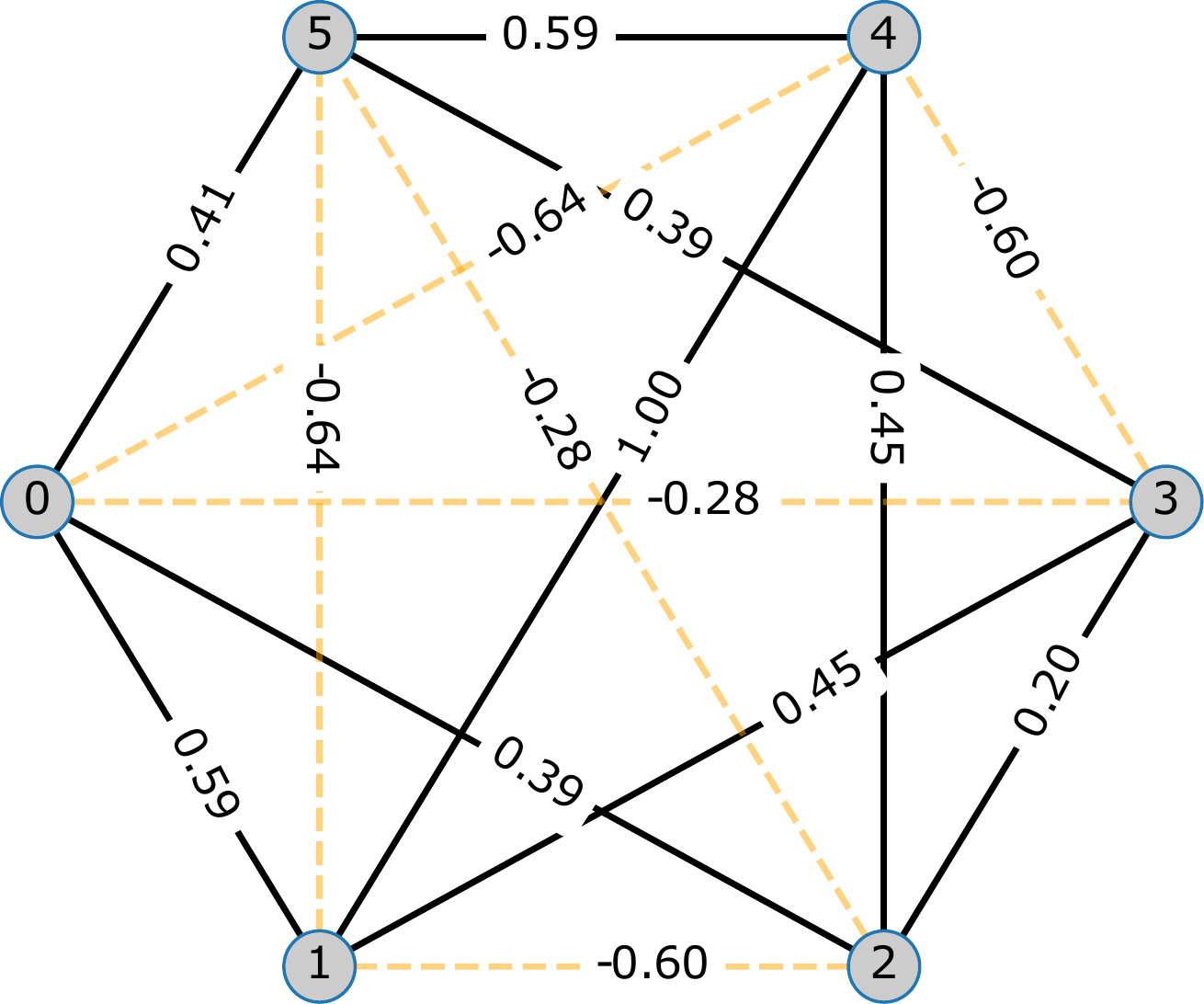}
     \caption{Ising interaction graphs for QAOA in Figs.~\ref{heatmaps 3 ion}-\ref{heatmaps 6 ion}, calculated with \refeq{Jij}.}
     \label{Cost instances}
\end{figure}

To apply the MS interaction to QAOA, it is important to choose experimental parameters that result in small final vibrational displacements, to minimize electron-vibration entanglement after the MS interaction. The experiments are detuned close to a target mode $m_t$ and we focus on conditions in which this mode is returned close to its initial state, since displacements of all other modes are small in the worst case.  This is accomplished by choosing interaction times $t = n_\mathrm{loops}t_\mathrm{loop}$, where $t_\mathrm{loop} = 2\pi/|\mu-\omega_{m_t}|$ is the time for the target vibrational mode to complete a single loop in phase space and $n_\mathrm{loops}$ is an integer.  Observed mode frequencies and chosen detunings are given in Table \ref{expt-freqs}, while the QAOA Ising couplings computed for these modes and detunings are visualized in Fig.~\ref{Cost instances}.  
\par
With a fixed set of gate times $t_\mathrm{gate} = n_\mathrm{loops}t_\mathrm{loop}$ in mind, to obtain two different QAOA parameters $\gamma$ and $\gamma'$ in Eq.~(\ref{gamma simple}) we need to scale the experimental parameters $\Omega$ and $t$ as
\begin{equation} \label{variable gamma} \frac{\gamma}{\gamma'} = \frac{\Omega^2 t}{\Omega'^2 t'},\end{equation} 
recalling that $J_\mathrm{max}\sim\Omega^2$. We fix an angle $\gamma_\mathrm{mp}$ as the angle that is achieved at maximum laser power in the experiments when $n_\mathrm{loops}=1$, with a corresponding Rabi rate $\Omega_\mathrm{mp}$ defining a $J_\mathrm{max}$ in Eq.~(\ref{gamma simple}).  To generate a different $\gamma$, we choose a number of loops
\begin{equation} n_\mathrm{loops} = \left\lceil \frac{\gamma}{\gamma_\mathrm{mp}} \right\rceil \end{equation}
where $\lceil \ldots \rceil$ is the ceiling function. We then scale the Rabi rate to obtain the desired $\gamma$,
\begin{equation} \Omega(\gamma) = \Omega_\mathrm{mp}\sqrt{\frac{\gamma}{n_\mathrm{loops}\gamma_\mathrm{mp}}} .  \end{equation}
Experimentally, the scaling in $\Omega$ is achieved by varying the laser power $P$. To apply the above relations, we need a calibrated value for $\gamma_\mathrm{mp}$ and $\Omega_\mathrm{mp}$; this is discussed in Appendix \ref{Rabi rate appendix}. 
\par  
\section{Calibrating and computing the Rabi rate} \label{Rabi rate appendix}
The Rabi rate $\Omega$ was not calibrated directly in the experiments.  Instead, we tuned the laser power to produce desired results in each experiment. This implies certain values for $\Omega$ that we computed as described below.
\par
In the experiments, we fixed a value for $n_\mathrm{loops}$ then varied the laser power to maximize an observable.  The observable that was maximized depends on the context.  For MS characterization experiments, we maximized the transition probabilities between strongly coupled states, $\vert 00\rangle \leftrightarrow \vert 11\rangle$. For QAOA, we either maximized transition probabilities or the expectation value of a simplified version of the cost function, assuming the $J_{i,j}$ only contain the dominant contributions from the target mode.  
\par
With the experimental procedures for setting the Rabi rate in mind, we now describe how to infer the values of $\Omega$ for our numerical calculations. For the MS results with two-ions in Fig.~\ref{2ion MS}, the experiments are designed to create a Bell state $\sim \vert 00\rangle -i\vert 11\rangle$ at time $t=3t_\mathrm{loop}$.  This corresponds to a factor $\chi_{0,1} = \pi/4$.  In the Ising approximation to the MS interaction Eq.~(\ref{Jij}), this implies that $\Omega/2\pi = \sqrt{\chi_{0,1}/[\sum_m \eta_{0,m}\eta_{1,m}\omega_mt/(\mu^2-\omega_m^2)]}/2\pi = 26.552$ kHz and we use this value in our computations.  
\par
For QAOA, we need the Rabi rate at max power $\Omega_\mathrm{mp}$, as defined in the $\gamma$ scaling procedure of Appendix \ref{gamma-setting}.  For this, we use noiseless numerical simulations to determine the QAOA angle $\gamma^*$ that maximizes a certain observable.  For three ions we maximized the population in $\vert 101\rangle$ after an MS application to $\vert 000\rangle$. For six ions there is no dominant basis state transition to maximize in the MS dynamics, so we instead ran QAOA and maximized the cost expectation at $\beta = \beta^*$, considering only the targeted mode in $J_{i,j}$. We tune the laser power in the experiment to maximize the previously stated quantities with fixed interaction times $t=n_\mathrm{loops}t_\mathrm{loop}$.  We set $\gamma_\mathrm{mp} = \gamma^*/n_\mathrm{loops}$ as the angle achieved in a single loop at this laser power and compute $\Omega_\mathrm{mp}$ from (\ref{gamma simple}). This yields $\Omega_\mathrm{mp}/2\pi=26.907$ kHz for three-ion QAOA and $\Omega_\mathrm{mp}/2\pi=27.690$ kHz with six ions.
\section{Finite sampling}\label{finite sampling}
Theoretical probabilities $P(z) = \langle z \vert \rho_I \vert z \rangle$ and cost expectations $\langle C \rangle = \mathrm{Tr}(\rho_IC)$ correspond to values that would be observed in an infinite set of measurements.  Experiments estimate these values using a fixed number of measurement shots $S$ as noted in the captions of Figs.~\ref{2ion MS}-\ref{heatmaps 6 ion}. This finite sampling alone leads to deviations from theoretical values computed as above, and we estimate the expected variation based on the standard error of the mean computed from $\rho_I$. This quantifies disagreement between the experimentally determined mean and the expected theoretical value. 
\par
For the MS characterization experiment in Fig.~\ref{2ion MS}, the probability estimated from $S$ measurements is a random variable $P_\mathrm{est}(z) = S_z/S$, with $S_z$ the number of times $\vert z \rangle$ was measured; its theoretical mean is $P_\mathrm{est}^\mathrm{mean}(z) = P(z) = \langle z \vert \rho_I\vert z\rangle$ with a standard error of the mean $\Delta P_\mathrm{est}=\sqrt{P(z)(1-P(z))/S}$ that quantifies the expected error from finite sampling, shown by colored bands in Fig.~\ref{2ion MS}. If experimental results are consistent with theory, then from the central limit theorem we expect about 2/3 of the experimental probabilities to be within $\Delta P_\mathrm{est}$ from the theoretical $P(z)$.  We use $\Delta P_\mathrm{est}^2$ as the variance in computing $\chi_\mathrm{red}^2$. 
\par
For the QAOA experiments in Figs.~\ref{heatmaps 3 ion}-\ref{heatmaps 6 ion} the standard error of the mean $\langle C \rangle = \mathrm{Tr}(\rho_IC)$ is $\Delta C = \sqrt{(\langle C^2\rangle - \langle C\rangle^2)/S}$.  We use this to compute standard errors of the approximation ratio in Eq.~(\ref{approximation ratio}) and for comparison of theoretical and experimental results in Table \ref{performance table}.
\section{SPAM measurements} \label{SPAM appendix}
\begin{figure*}
    \centering
    \includegraphics[width=0.9\textwidth]{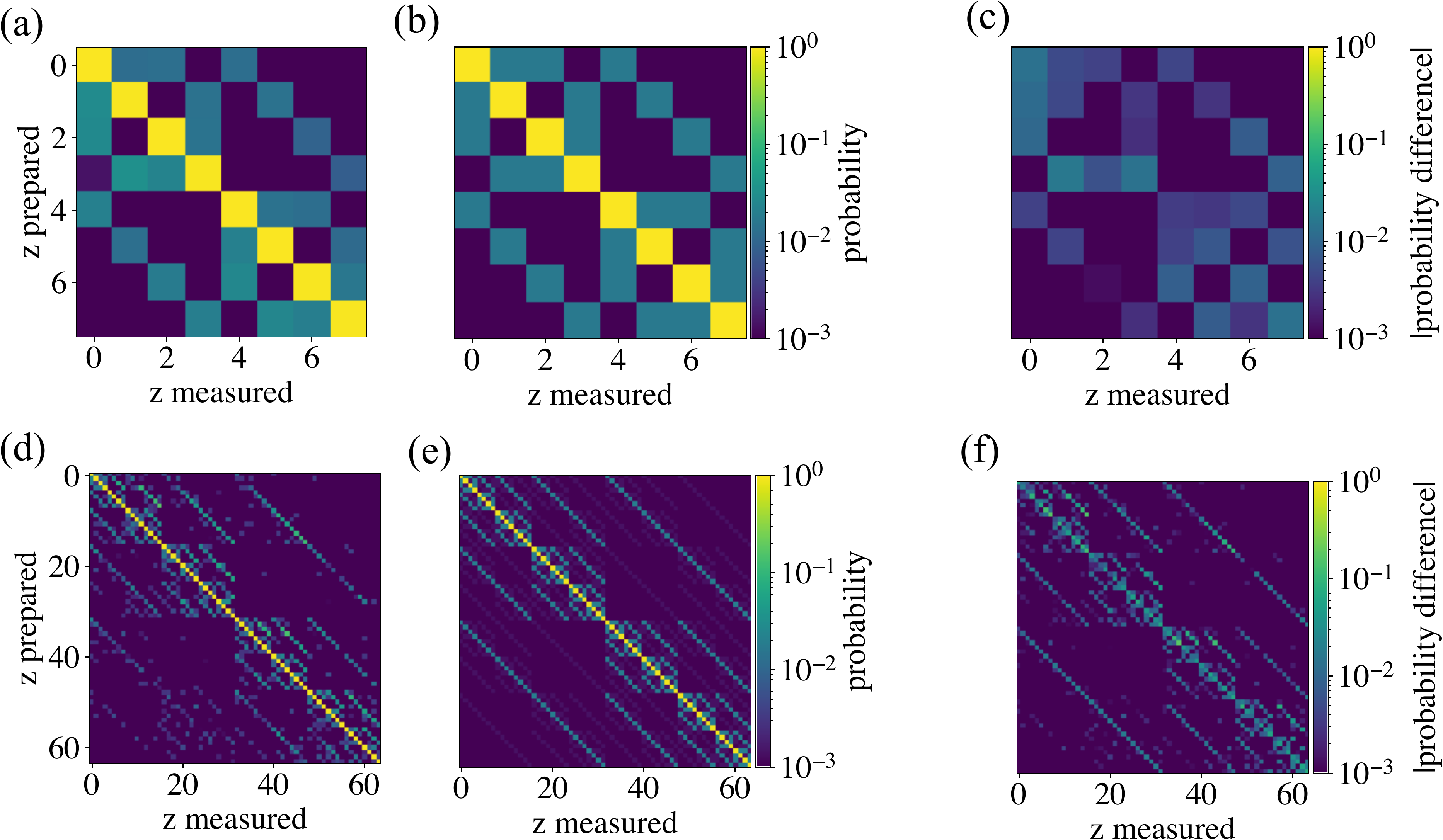}
    \caption{SPAM matrices in experiment and theory. (a) Experimental 3-ion SPAM matrix with $S=4000$ shots per prepared state, (b) best fit SPAM matrix with independent bit-flip errors, (c) the difference between them.  (d) Experimental 6-ion SPAM matrix with $S=500$ shots per prepared state, (e) theory with bit flip errors $2\%$, and (f) the difference.}
    \label{SPAM}
\end{figure*}

To measure SPAM errors experimentally, we prepared all possible $Z$-eigenstates and measured the distribution of states observed. The experiment is equipped with a pair of tightly focused laser beams, split from the same mode-locked 355-nm laser used to generate global Raman gates. With a $1/e^2$ intensity radius of 4.5(1)~$\mu$m, 18~mW gives an individually addressed $Z$-rotation with a 66~$\mu$s pi-time due to a 4th-order light shift \cite{lee_engineering_2016}. Every ion can be individually addressed in a single experiment through adiabatic transport of the entire chain similar to Ref.~\cite{Herold2016}. Combined with global $R_Y(\pm \pi/2)$ rotations, we transform the addressed $Z$-rotations into bit flips to prepare any $Z$-eigenstate. The observed SPAM matrices are shown in Figure \ref{SPAM}, and are expected to be dominated by measurement errors as addressing bit flip errors are $<10^{-3}$ for the three-ion chain and $<10^{-2}$ for the six-ion chain.
\par
The SPAM errors can be approximately accounted for using a model of independent bit flips in each ion. In this model the SPAM matrix $M_{z_I',z_I} = \epsilon^{h(z_I',z_I)}(1-\epsilon)^{n-h(z_I',z_I)}$, where $h(z_I',z_I)$ is the Hamming distance between $z_I'$ and $z_I$ and $\epsilon=0.02$ is an effective error probability, based on a fit that minimizes the trace distance between experimental and model SPAM matrices at three and six ions, see Table \ref{tab:SPAM}.  The best-fit model SPAM matrices are also pictured in Fig.~\ref{SPAM} along with the residual error compared to experiment; the best-fit error probability is $\epsilon=0.02$. Similar errors are found by minimizing the sum of absolute differences of the individual matrix elements.
\begin{table}[H]
    \caption{The trace distance between experimental and theoretical SPAM matrices, where theory is calculated without error or with a 2\% bit-flip error in each qubit.}
    \label{tab:SPAM}
    \centering
    \begin{ruledtabular}
    \begin{tabular}{ccc}
        $n$ & Trace distance no error & Trace distance 2\% error \\
        \hline
        3 & 0.21 & 0.04\\
        \hline
        6 & 3.68 & 1.44\\
    \end{tabular}
    \end{ruledtabular}
\end{table}
\bibliography{references}
\end{document}